\documentstyle[aps]{revtex}

\renewcommand{\baselinestretch}{1.7}
\begin{document}
\draft
\preprint{IP/BBSR/94-23}
\preprint{06 May 1994}
\title {   Magnetic Wormholes and Vertex Operators }
\author{ Harvendra Singh\cite{mail} }
\address{Institute of Physics, Bhubaneswar-751 005, INDIA.}
\maketitle
\begin{center}
\end{center}
\begin{abstract}

We consider wormhole solutions in $2+1$ Euclidean dimensions. A
duality transformation is introduced to derive a new action from
magnetic wormhole action of Gupta, Hughes, Preskill and Wise. The
classical solution is presented. The vertex operators corresponding
to the wormhole are derived. Conformally coupled scalars and spinors
are considered in the wormhole background and the vertex operators are
computed.
\end{abstract}

\narrowtext

\newpage

\section {Introduction}
\label{int}

\par Recently, considerable attention has been focused on the
study of topology changing processes. The effects of wormholes are
interesting and they play an important role in  the quantum theory of
gravity. Hawking~\cite{haw1} has argued that such processes might be
responsible for the loss of quantum coherence. Coleman has advanced
the proposition that wormholes introduce indeterminacy in the
constants of Nature~\cite{col2}. He has argued persuasively that the
constants of Nature are randomly distributed in the ensemble of
universes. Furthermore, he argued that the probability distribution
is peaked for zero value of the cosmological constant~\cite{col1}.
There has been considerable amount of activity following the work of
Coleman some of which are given in~\cite{pre}. However, the
mechanism proposed by Coleman has been subjected to criticism. It is worthwhile
to ask how much the physics at low energies is affected due to the
wormholes. Hawking~\cite{haw1} has introduced the concept of vertex
operators to account for the effect of wormholes at energies much
smaller than the Plank  scale. Thus it is possible to construct
effective Lagrangians at low energies.
\par The purpose of this note is to study wormholes in $2+1$
euclidean dimensions. We consider the Lagrangian introduced by Gupta
et al.~\cite {gup-hug-pre-wis} where abelian gauge field is
responsible for the existence of classical solution of the
Einstein-Maxwell equation. This is the so called the magnetic wormhole
solution. Recently, it has been shown ~\cite{kha-mah} that such
3-dimensional action can be derived from a 4-dimensional theory by
adopting dimensional reduction technique. The 4-dimensional theory
has Einstein-Hilbert action and the action of an antisymmetric
tensor field. The dimensionally reduced action , suitably choosing
the compactification, gives rise to the action of Gupta et al.
\par We introduce a duality transformation so that a scalar field is
dual to the electromagnetic field strength. The field equations are
presented and explicit solutions are given for the geometry
$R^1\,\times \,S^2 \,$.
\par It is observed that there is a global conserved charge in this
theory which is responsible for stabilising the classical wormhole
solution . This property is very useful to construct vertex operators
for the wormholes. Recently, Hawking~\cite{haw3} and
Grinstein and Maharana~\cite{gri-mah} have analysed the structure of
these vertex operators explicitely for the axionic wormholes.
The effects of wormholes of size
smaller than the scale of the observation are taken care of once the
gauge-invarient, bi-local effective interaction terms, vertex
functions, are added to the original action.
Presently, it is widely beleived that the considerable knowledge
about these vertex operators may provide a good understanding of the
topology change in quantum gravity.
\par Next we consider matter fields in the background of the
wormhole. These fields are not responsible for the existence of the
classical wormhole solutions. However, the correlation functions for
these field configurations, while they are far away from the throat
of the wormholes, are expected to be different from flat space
correlation functions. We compute such correlation functions by
introducing the vertex operators. The vertex operators for the scalar
fields and spinor fields are presented in section-IV . The last
section-V contains summary of our results and discussions.

\section{ 2+1-Dimensional Magnetic Wormhole }
\label{magnetic}
\par The magnetic wormhole solutions in three euclidean dimensions
 have been obtained by Gupta et al.~\cite {gup-hug-pre-wis}. The action is

\begin{equation}
S= \int d^3x \sqrt{g}\left (-{R\over 16\pi G}+{1\over 4 e^2}F^2 \right )
\label{21}
\end{equation}

\noindent where $G=(16\pi)^{-1}M_P^2$ and $e$ are newtonian gravitation
constant and electromagnetic coupling constant, respectively. $M_P$ is the
Plank mass and $\hbar=c=1$. Here the vector
potential is such that the field strength admits a   monopole like
configuration . The Einstein gravitational
field equations are

\begin{equation}
G_{\mu\nu}={M_P^2\over 4e^2}\left
(2F_{\mu\lambda} F_{\nu}^\lambda-{1\over 2}g_{\mu\nu}F^2\right )
\label{22}
\end{equation}
\noindent where

\begin{equation}
G_{\mu\nu}=R_{\mu\nu}-{1\over 2}g_{\mu\nu}R\,\,\, .
\nonumber
\end{equation}
\noindent While the matter field equation is

\begin{equation}
{1\over\sqrt g} \partial_\mu\left(F^{\mu\nu}\sqrt{g}\right)=0\,.
\label{23}
\end{equation}

\noindent Metric and the field strength, which satisfy  the above
field equations, have the following form

\begin{equation}
ds^2= dt^2 + a^2(t) \left (d\theta^2+sin^2\theta d\phi^2 \right)
\label{24}
\end{equation}

\begin{equation}
F= f(t) \varepsilon\\
 =f(t) a^2(t) d\theta \wedge sin\theta d\phi \, .
\label{25}
\end{equation}

\noindent Integral over $\varepsilon$  gives the surface area of the two
sphere,
$\int_{S^2}\varepsilon=4\pi a^2(t)$.

\par The  magnetic flux is given by
\begin{equation}
\Phi=\int_{S^2}F =\int _{S^2} F_{\mu\nu} d\Sigma^{\mu\nu} =n \Phi_o
\label{26}
\end{equation}
\noindent and thus
\begin{equation}
f(t)={n \over 2 a^2 (t)}\,\, .
\label{27}
\end{equation}
\par The magnetic flux , $\Phi$, is  required to be the integer multiple of
unit quantum flux
$\Phi_{o}$ ( equal to $2\pi$) if particles of the unit charge are to
be introduced on the wormhole background. The scale factor, $a(t)$, satisfies
the equation (from $o-o$ component of the Einstein equation)

\begin{equation}
(\partial_t a)^2 = 1- {a^2_o \over a^2} \,\,,
\nonumber
\end{equation}
\noindent where
\begin{equation}
 a^2_o={\pi\, G \,\,n^2\over e^2}
\nonumber
\end{equation}

\noindent and solution to $a(t)$ is

\begin{equation}
a^2=a^2_o +t^2\,\,.
\label{28}
\end{equation}

Thus the wormhole configuration is such that there are two
asymptotically  flat regions ( corresponding to $t=\underline{+}
\infty$ ) connected by a tube of minimum throat size  $a_o \>$.
\par It is convenient to introduce a dual field of electromagnetic
field strength $ F_{\mu\nu}$ . We may recall in this context the four
dimensional axionic wormhole solutions described by Giddings and
Strominger~\cite{gid-str1} and Myers~\cite{mye}. the effective
action of string theory contains the square of the field strength of
the antisymmetric field strength tensor $ B_{\mu\nu}$. The effective
action considered in the ref.~\cite{gid-str1} is

\begin{equation}
\int \left (-{1\over 16\pi G } R +{1\over
12}H_{\mu\nu\lambda}\,H^{\mu\nu\lambda}\right )\sqrt{g} d^4x \,\,.
\nonumber\end{equation}
The field strength can be locally expressed as
\begin{equation}
H_{\mu\nu\lambda}=\varepsilon_{\mu\nu\lambda\rho}\,\,\partial^\rho \cal{A}
\nonumber
\end{equation}
\noindent where $\cal {A} $ is pseudoscalar field, so that
$d^\ast H=0=\,dH$. Here asterisk
denotes the Hodge dual. In three euclidean
dimensions, we can write

\begin{equation}
F_{\mu\nu}= \varepsilon_{\mu\nu \lambda} \nabla^\lambda \phi
\label{29}
\end{equation}
or $^\ast F= d\phi$ is a closed one-form; $d^\ast F=0=\,dF$.  The dual action
is obtained by substituting
eq.(\ref{29}) in the eq.(\ref{21}),
\begin{equation}
S=\int d^3x \sqrt{g} \left (-M_P^2 R+{1\over 2e^2}(\nabla\phi)^2 \right)\,\,.
\label{210}
\end{equation}
It describes a theory of the scalar field coupled to gravity. This
scalar field is directly responsible for the stability of the
wormhole and providing it with negative stress energy.
For the spherically symmetric ansatz, eq.(\ref{24}), of the metric
the Einstein and matter field equations are satisfied for

\begin{equation}
\phi={n\over 2 a_o} \arctan{t\over a_o}\,\,.
\label{211}
\end{equation}
We note that the field equation for the dual scalar field, $\phi$,
has the form of a current conservation law and consequently

\begin{equation}
\int_{S^2}\left (\sqrt{g} g^{o o}\partial_o\phi \right )= 2\pi n\,,
\label{212}
\end{equation}
where the integration is taken over 2-sphere around t=0. Here n is
required to be integer after quantisation.

\section{ Vertex operator}
\label{ver}
\par The concept of vertex operator, in the context of wormholes, was
first introduced by Hawking~\cite{haw1}. In order to account for the effects of
wormholes on the physical phenomena, at length scales much larger
than the wormhole size, one can construct effective actions following
the proposal of Hawking. We can envisage the scenario where the
initial configuration is characterized by the geometry $\Sigma_i$
with charge $Q_i$ which evolves to a geometry $\Sigma_f$ and
charge $Q_f$ . The amplitude can be represented in the Feynman's
path integral approach. However, it must be emphasized that the
wormhole is stabilized by a conserved global charge,

\begin{equation}
Q=\int_\Sigma d\Sigma^\mu J_\mu
\label{31}
 \end{equation}
which follows from the current conservation equation, ${J^\mu}_{;\,\mu}=0$.
An elegant formulation of the topology changing processes, in
Hamiltonian path integral approach, was given by Grinstein~{\cite{gri}}.
In what follows we adopt the ref. {\cite{gri}}.
 \par Let us first write the transition amplitude from an arbitrary
initial state
$\mid \Sigma_i,\phi_i >$ to some final state $\mid \Sigma_f,\phi_f >$
which in path integral representation is
given by ~\cite{haw4},

\begin{equation}
\langle \Sigma_f,\phi_f \mid \Sigma_i,\phi_i \rangle=\int_{b.c.} D[g,\phi]
\exp\left(-S(g,\phi)\right)
\label{32}
 \end{equation}
 where $D[g,\phi]$ is a measure on the space of all field
configurations $g$ and $\phi$, $S( g,\phi) $ is the action of the
fields, and the integral is taken over all fields which have given
boundary conditions $(b.c.)$; $\phi(x, t=t_f)= \phi_f$ and $\phi( x,t=t_i)=
\phi_i$ on
surfaces $\Sigma_f$ and $\Sigma_i$
respectively. We are interested in initial and final configurations
with definite charges. Such states are defined by geometries
$\Sigma_i$,$\,\,\Sigma_f$ and charge densities $J_\phi^i$,
$J_\phi^f$ respectively. Here $J_\phi$ , the charge density, is the canonically
conjugate momentum of the scalar field $\phi$. After some lengthy but
straight forward calculations, we arrive at the following expression
for the transition amplitude,
\begin{eqnarray}
\langle \Sigma_f,J_\phi^f\mid\Sigma_i,J_\phi^i\rangle&=&\int d\phi_i
d\phi_f{e^{i [\int_{\Sigma_f} d\Sigma\cdot J_\phi^f \phi_f
-\int_{\Sigma_i}d\Sigma\cdot J_\phi^i \phi_i]}}\times\langle \Sigma_f,\phi_f
\vert\Sigma_i,\phi_i\rangle \nonumber \\
\nobreak &=&\int d\phi_i d\phi_f e^{i[Q_f\phi_f -Q_i \phi_i]} \int D[g,\phi]
\exp {(-S[g,\phi])} \nonumber \\
& &
\label{33}
\end{eqnarray}
where $Q_i=\int_{\Sigma_i} d\Sigma\cdot J_\phi^i$ and $Q_f=\int_{\Sigma_f}
d\Sigma\cdot J_\phi^f$ are the conserved charges on
respective surfaces. Using the identity

\begin{eqnarray}
  \phi_f=\int \limits_{t_i}^{t_f} dt \partial_t \phi + \phi_i
\nonumber
\end{eqnarray}
we can write
\begin{eqnarray}
[Q_f \phi_f -Q_i \phi_i] = i[Q_f-Q_i] \phi_i +i{\int\limits_{t_i}^{t_f} dt
\int_{\Sigma_f}
d\Sigma_\mu J_\phi ^\mu \partial_t\phi}\,\,.
\nonumber
\end{eqnarray}
Finally the integration over $\phi_i$ and $\phi_f$ will give
\begin{equation}
\langle\Sigma_f,J_\phi^f\vert\Sigma_i, J_\phi^i\rangle= \delta
(Q_f-Q_i) \int D[g,\phi] {e^{-S + i \int \limits_{t_i}^{t_f}
d^3x \sqrt{g} J_\phi^\mu\partial_\mu\phi}}
\label{34}
\end{equation}
where we have used the fact that $\partial_i\phi=0$, i.e., $\phi$
does not vary over space like surface $\Sigma$, see eq.(\ref{211}). The above
expression tells us that charge of initial and final configurations
should be conserved. Now we are ready to do integration over $\phi$.
The result of the integration of $\phi$ is reproduced if we set into the
integrand
on the right hand side of the above equation
\begin{equation}
\phi^b_{;\, \mu} = iJ_\mu
\label{35}
 \end{equation}
and remove the integration over $\phi$. The value of the field, $\phi^b$, in
above equation corresponds to the effective background or saddle
point configuration. Note that we have dropped suffix $\phi$ of
$J_\phi ^\mu$ and here onward we shall write only $J^\mu$.
\par The path integral in eq.(\ref{34}) does include the
wormhole fluctuations. It is not known so far how to handle
these fluctuations in a spacetime theory explicitly. However, it
does not stop one to formulate a low energy effective theory.
Here one can get rid of the fluctuations of wavelengths shorter
than some scale $L\gg a_o$ by integrating them out. In this
approach the effects of wormhole
fluctuations of size smaller than $ L$ can be summarized through inserting
$\it vertex \,\, functions$ in the integrand~\cite{haw1}. The
transition amplitude can be written in the following
factorized form by introducing the vertex operators,

\begin{equation}
\langle\Sigma_f, J_\mu^f\vert \Sigma_i, J_\mu^i\rangle
=\langle\Sigma_f,J_\mu^f\vert V\rangle_o
\times \langle V \vert\Sigma_i,J_\mu^i\rangle_o
\label{35a}
\end{equation}
where
\begin{equation}
\langle\Sigma, J_\mu\vert V\rangle_o=\int D[g,\phi] {e^{-S + i\int
d^3x\sqrt{g} (J^\mu \phi )_{;\, \mu}}}\,\,V \,\,\,.
\label{36}
\end{equation}

Note that the amplitudes on
the right hand side of (\ref{35a}) are computed in flat space
and integral over configurations
does no more include wormhole fluctuations. Their sole effect is
accounted by the introduction of the vertex operator, $ V\,\,$. Further
wormhole carries
charge and one should no more insist on $J^\mu_{;\,\mu}=0$ rather following is
true,

\begin{equation}
J^\mu_{; \mu}= Q{\delta^3(x-x_o)\over\sqrt g},
\label{37}
\end{equation}
and
\begin{eqnarray}
\int d^3x\sqrt {g}( J^\mu\phi)_{;\, \mu}= Q \phi(x_o) + \int d^3x \sqrt{g}
J^\mu \phi_{;\,\mu}\,\,\,.
\nonumber
 \end{eqnarray}

\par It  determines the vertex operator, $V$, on the r.h.s. of the
eq.(\ref{36}) to be
$\sim e^{-iQ \phi(x_{o})}$ for compensation.
\par We present below an alternative  method of extracting vertex operators
{}~\cite{gri-mah}. Here expectation value of field
operators in wormhole background is approximated by their saddle
point values in semiclasical limit$ (\hbar=0)$,

\begin{equation}
\langle\phi(x_1)\cdots\phi(x_n)\rangle_w=\langle1\rangle_w
\phi^b(x_1)\cdots\phi^b(x_n) + O(\hbar)
\label{38}
\end{equation}
where $< 1 >_w$ is the normalisation and  $\phi^b(x)$ is the saddle
point value of the scalar field. We can read  from equations (\ref{35})
and (\ref{37}) that $\phi^b$ satisfies the green function equation,
\begin{equation}
\Box\,\phi^b(x)= iQ\,{\delta^3(x-x_o)\over\sqrt g}\,\,.
\label{39}
\end{equation}
Eq. (\ref{39}) is the same as the one satisfied by  the Feynman propagator,
$\bigtriangleup_{F}
(x,x_{o})\,$. Thus we can write the product of fields on right hand side of
eq. (\ref{38}) as
\begin{equation}
\langle\phi(x_1)\cdots\phi(x_n)\rangle_w= \langle1\rangle_w
(-i\,Q)^n \prod\limits_{i=1}^n \bigtriangleup_F (x_i,x_o)\,\,.
\label{310}
\end{equation}
In order to reproduce the above results, we must chose an appropriate
form for the vertex operator. It is evident that $V(\phi(x_o))$
should have the following form

\begin{equation}
V(x_0)=\langle 1 \rangle_w {1\over n! }(-i\,Q)^n \,\phi^n(x_o)
\label{311}
\end{equation}
so that relation

\begin{equation}
\langle
\phi(x_1)\cdots\phi(x_n)\phi(y_1)\cdots\phi(y_n)\rangle_w=\langle
\phi(x_1)\cdots\phi(x_n)\,
V(x_o)\rangle_o  \langle V(\tilde{x}_o) \phi(y_1)\cdots\phi(y_n) \rangle_o
\label{312}
\end{equation}
\noindent can be checked to be consistent with eq.(\ref{38}) when we
choose  $V$ as given by (\ref{311}). Note that points
$\{x_1,\cdots,x_n \}$ belong to one asymptotically flat region
whereas $\{ y_1, \cdots ,y_n\}$ belong to the another asymptotically
flat region. The tilde in $V(\tilde{x}_o)$ is used to
distinguise the vertex operators in two
asymptotic regions of the wormhole. We have obtained the form of the
vertex operators for product of a string of n fields at different
spacetime points. The general expression for $ V$ now can be
written
\begin{equation}
V(x_o)= \langle1\rangle_w\, e^{-i Q\phi(x_o)}\,\,\,.
\label{313}
\end{equation}
\par Now we turn our attention towards the emission and
absorption of gravitons by the wormholes. It is expected that one
might be able to compute correlation functions inolving the metrics
$g_{\mu \nu}(x_i)$. Such correlation function  will be gauge
dependent. One can sideline this difficulty by
calculating correlations involving the object like $\prod\limits_{i}
R(x_i)\,\,$, where $R(x)$ is the scalar curvature. The result is
given by the background configuration,
\begin{equation}
\langle R(x_1)\cdots R(x_n)\rangle_w=\langle1\rangle_w R^b(x_1)
\cdots R^b(x_n)\, + O({\hbar})
\label{323}
\end{equation}
where background curvature is ,
\begin{equation}
R^b(x)\,=-{8\,r^2_o \over \vert x -x_o\vert^4 {(1+{r^2_o\over\vert
x-x_o\vert^2})^4}}.
\label{324}
\end{equation}
Since the right hand side of eq.(\ref{323}) involves products of
curvature scalars at their saddle point values thus we are entitled
to use the field equation to facilitate semiclassical computations.
One can check that the vertex operators derived for the fields,
$\phi$, correctly reproduce the desired results when we use the
relations. Thus

\begin{eqnarray}
\langle R(x_1)\cdots R(x_n)\rangle_w=(8\pi\,G)^n\langle
\phi^{,\mu}\phi_{,\mu}(x_1)\cdots\phi^{,\mu}\phi_{,\mu}(x_n)\,V(x_o)\rangle_o
\langle V(\tilde{x}_o)\rangle_o + O(\hbar)\,.
\nonumber
\end{eqnarray}
\par  In the following section we shall also include matter fields in wormhole
background.
  Correspondingly the structure of vertex operators will get modified.

\section{ Matter Field Sector}
\label{field-sector}
\par First let us write the metric of Gupta et
al.~\cite{gup-hug-pre-wis}, eq.(\ref{24}), in the following
conformally flat form

\begin{equation}
ds^2= \Omega^2(x) \delta_{\mu\nu} dx^{\mu}\;dx^{\nu}
\label{41}
\end{equation}
where
\begin{equation}
\Omega^2(x)=\, \left (1+ {{a^2_o\over 4}\over \vert x-x_o\vert^2} \right)^2
\label{42}
\end{equation}
and we find
\begin{equation}
\vert x-x_o\vert={a_o\over2} \exp (\sinh^{-1}{t\over a_o})\,\,.
\label{43}
\end{equation}
\par We note that two asymptotically flat regions correspond to
$x\rightarrow\infty$(or $t\rightarrow\infty$) and $x\rightarrow
x_o$(or $t\rightarrow -\infty$). The singularity at
$x=x_o$ is not a curvature singularity. It can be seen from the
inversion of the coordinates in the sphere of radius $a_o/2$ as
follows
\begin{equation}
(y-y_o)_\mu (x-x_o)^\mu={a^2_o\over 4}\,=r^2_o \;(say)\,\,.
\label{44}
\end{equation}
This brings the point at infinity and the null infinity surface in
the x-space to the origin and the light cone at the origin in
y-space, respectively. The origin of the original spacetime(x-sace)
and its light cone are sent to infinity. Metric in y-space becomes
conformal to that of x-space
\begin{eqnarray}
dy^\mu\,dy_\mu=\left ({r^2_o\over \mid x-x_o
\mid^2}\right)^2\,dx^\mu\,dx_\mu \,\,,
\nonumber
\end{eqnarray}
whence eq.(\ref{41}) becomes
\begin{eqnarray}
ds^2=\left(1+{ \mid x-x_o \mid ^2\over r^2_o}\right)^2 \, dy^\mu\,dy_\mu\,.
\nonumber
\end{eqnarray}
\par Now we are ready to find correlation functions for matter fields
in wormhole background. Though this, generally, is a difficult  task
. But the calculation is much simplified for conformally coupled
fields in conformal backgrounds. These fields have zero vacuum
expectation values.

\par We devide the spacetime region of the wormhole into two subspaces,
viz., $\vert x_i-x_o\vert\gg r_o$ and $\vert x_i-x_o\vert\ll r_o$. We
shall denote later ones by a tilde. Let us first consider the propagation
of the scalar fields , $S(x)$ , from one side of the wormhole to the other
side. Since scalar field is conformally coupled in three dimensions,
it immediately gives for two-point correlation function in wormhole
background
 \begin{equation}
 \langle S(x)\,S(\tilde{x})\rangle_w=\langle1\rangle_w
\Omega^{-1/2}(x)\,\Omega^{-1/2}(\tilde{x})\bigtriangleup
_F(x,\tilde{x})
\label{45}
\end{equation}
where the flat space propagator is
\begin{equation}
 \bigtriangleup _F (x,\tilde{x})=
 \langle S(x)\,S(\tilde{x})\rangle_o={1\over 4\pi}{1\over\vert
x-\tilde{x}\vert}\,\,.
\label{46}
\end{equation}
In the limit $\mid x-x_o\mid\gg r_o\gg\mid\tilde{x}-x_o\mid$
\begin{equation}
\Omega(x)\approx1 ,\;\;\;
\;\Omega(\tilde{x})\approx{r^2_o\over\mid\tilde{x}-x_o\mid^2},\\
\,\,\, \mid x-\tilde{x}\mid\approx\mid x-x_o\mid.
\label{47}
\end{equation}
We get to leading order in $\,r_o\over\vert x-x_o \vert\,$ and
$\,\vert\tilde{x}-x_o\vert \over r_o \,$,
\begin{equation}
\langle S(x)\,S(\tilde{x}\rangle_w\approx\langle1\rangle_w \left
({r_o\over\mid\tilde{x}
- x_o \mid}\right )^{-1}{1\over 4\pi}{1\over \mid x-x_o\mid}\,\,.
\label{48}
\end{equation}
Now, we can evaluate the
vertex operators once we write the first factor in the inverted
coordinates defined in eq.(\ref{44}). In these coordinates the
distance from the wormhole throat at $x_o$ to the point $\tilde{x}\,\,$
,$\,\approx {r_o^2 \over\vert \tilde{x}-x_o \vert}\,\,$, becomes $\,\vert
\tilde{y}-\tilde{y}_o\vert$. Eq.(\ref{48}) can be expressed as
\begin{equation}
\langle S(x)\,S(\tilde{y}\rangle_w\approx\langle1\rangle_w r_o{1\over
\mid\tilde{y}
-\tilde{y}_o \mid}{1\over 4\pi}{1\over \mid x-x_o\mid}\,\,.
\label{49}
\end{equation}
If one  looks carefully, it is clear that the above correlation
function is in the factorised form. Let us write
$\langle S(x)\, S(\tilde{y})\rangle_w$ in the factorised form by
introducing appropriate vertex operators
 \begin{equation}
\langle S(x)\,S(\tilde{y})\rangle_w=\langle S(x)\,V(x_o)\rangle_o\langle
S(\tilde{y})\,V(\tilde{y}_o)\rangle_o
\label{410}
\end{equation}
Thus it follows from equations (\ref{313}), (\ref{49}) and
(\ref{410}) that the vertex operators associated with the two point
function of conformally coupled scalar field is
\begin{equation}
V(x_o)\,V(\tilde{y}_o)= c^2 e^{-i\,Q\,\phi(x_o)
+i\,Q\,\phi(\tilde{y}_o)} 4\pi r_o S(x_o)\,S(\tilde{y}_o)\,\,.
\label{411}
\end{equation}
Note that this factorization of the correlation function in the
wormhole background is
 crucial for the determination of the vertex
operators. The normalisation constant $c^2={\langle1\rangle_w\over
\langle1\rangle_o^2}$ corresponds to the ratio of unit operators
evaluated in wormhole background and in flat space.
\par The above results are easily extendible to the case when $n$
fields are inserted on each side of the wormhole, i.e.,
\begin{equation}
\langle\prod\limits_{i=1}^n
S(x_i)\,S(\tilde{y}_i)\rangle_w=\langle1\rangle_w\,(4\pi\,r_o)^n n!
\prod\limits_{i=1}^n {1\over4\pi\mid x_i-x_o\mid}{1\over 4\pi\mid\tilde{y}_i
-\tilde{y}_o\mid}\,\,.
\label{412}
\end{equation}
This gives vertex function to be
\begin{equation}
V(x_o)\,V(\tilde{y}_o)={c^2 (4\pi\,r_o)^n\over
n!}e^{-iQ[\phi(x_o)-\phi (\tilde{y}_o)]} \,\,S^n(x_o)\,\,S^n(\tilde{y}_o)\,\,.
\label{413}
\end{equation}
A factor of $({1\over\hbar})^n$ is hidden on the right hand side of the above
expression for the vertex operator in order to get correct order
$\hbar^{2n}$ for $2n$ propagators. Whence together one can write a
complete expression for the vertex operator
\begin{equation}
V(x_o)\,V(\tilde{y}_o)=c^2
\,e^{-i\,Q\,[\phi(x_o)-\phi(\tilde{y}_o)]}
\,\,\exp[4\pi\,r_o\,S(x_o)\,\,S(\tilde{y}_o)]\,\,.
\label{414}
\end{equation}

\par Massless Fermi fields, too, are conformally scaled with the
conformal scaling of the metric, e.g.,  for metric given in eq.
(\ref{41}), we will have

\begin{equation}
\psi (x)\rightarrow \Omega^{-1}(x)\,\,\psi(x)\,.
\label{415}
\end{equation}
The propagator in wormhole background can be written as

\begin{equation}
\langle \psi (x) \, \bar{\psi}(\tilde{x})\rangle_w =\langle1\rangle_w
\Omega^{-1}(x)\Omega^{-1}(\tilde{x})\, S_o(x,\,\tilde{x})
\label{416}
\end{equation}
where $ S_o(x,\,\tilde{x})=\langle \psi (x) \,
\bar{\psi}(\tilde{x})\rangle_o =-{i\over 4\pi}{\gamma^a
(x-\tilde{x})_a\over \vert x-\tilde{x}\vert^3}\,\,$ which is the flat
space propagator for fermi fields. Here $\gamma^a \,$ 's are
$2\times2$ matrices with
$\,\{\gamma^a,\gamma^b\}=2\,\delta^{a\,b}\,I,\,\,a=1,2,3$, where $I$
is the identity matrix. We shall
find out vertex operators for Lorentz scalar operators, like
$\bar{\psi}(x)\,\psi(x)$ . One will get

\begin{equation}
\langle\bar{\psi}(x) \psi(x)\,\bar{\psi}(\tilde{x})
\psi(\tilde{x})\rangle_w= \langle1\rangle_w\,Tr
S_w(x,\,\tilde{x})\, S_w(\tilde{x},\, x)
\label{417}
\end{equation}
where $Tr$ represents the trace over spinor indices. On simplification we get

\begin{equation}
\langle\bar{\psi}(x)\psi(x)\,\bar{\psi}(\tilde{x})
\psi(\tilde{x})\rangle_w =-{\langle1\rangle_w \over
(4\pi)^2}\Omega^{-2}(x)\,\Omega^{-2}(\tilde{x}) \,{2\over\vert
x-\tilde{x}\vert^4} \,.
\label{418}
\end{equation}

Since $\vert x-x_o\vert \gg r_o \gg\vert \tilde{x}-x_o \vert\:$,
using eq.(\ref{47}) we obtain

\begin{equation}
\langle\bar{\psi}(x)\psi(x)\,\bar{\psi}(\tilde{x})
\psi(\tilde{x})\rangle_w \cong -{\langle1\rangle_w \over
(4\pi)^2}{\vert\tilde{x}-x_o \vert^4 \over r^4_o}{ 2\over \vert x-x_o
\vert^4}
\label{419}
\end{equation}
which indeed gets factorized if one notices eq.(\ref{44}).
This result is reproduced if we insert the  vertex operator

\begin{equation}
V(x_o)\,V(\tilde{x_o}) = c^2\,
e^{-i\,Q\,[\phi(x_o)-\phi(\tilde{x_o})]}\, (-8\pi^2\,r_o^4)
\bar{\psi}(x_o)\psi(x_o)\,\bar{\psi}(\tilde{x}_o)
\psi(\tilde{x}_o)
\label{420}
\end{equation}
in the following

\begin{equation}
\langle\bar{\psi}(x)\psi(x)\,\bar{\psi}(\tilde{x})
\psi(\tilde{x})\rangle_w =
\langle\bar{\psi}(x)\psi(x)\,V(x_o)\rangle_o \,
\langle\bar{\psi}(\tilde{x})\psi(\tilde{x})\, V(\tilde{x}_o)\rangle_o
\,.
\end{equation}
The above calculation can be done for arbitrary number of insertions of
operators $\bar{\psi}\,\psi$ on either side of wormhole. Whence we
obtain a complete expression for the vertex operator in case of
fermion propagation,
\begin{equation}
V(x_o)\,V(\tilde{x_o}) = c^2\,
e^{-i\,Q\,[\phi(x_o)-\phi(\tilde{x_o})]}\, \exp \left[(-8\pi^2\,r_o^4)
\bar{\psi}(x_o)\psi(x_o)\,\bar{\psi}(\tilde{x}_o)
\psi(\tilde{x}_o)\right]
\label{421}
\end{equation}

Equations (\ref{414}) and (\ref{421}) are the complete vertex
operators for the scalar and spinor fields evaluated in the
background of the wormhole. We note that the matter field parts of
these vertex operators are not factorized in the fields $S(x)$ and
$\psi(x)$ respectively. If we were introduced $\alpha$-parameters
following the formalism due to Coleman~\cite{col1}, it is evident that
 the $\alpha$-parameters will now be labeled by the global $Q$ and
 the attributes of the corresponding matter fields.

\section{Conclusion and Discussion}

\par We have worked out the dual theory for magnetic wormhole of
Gupta et al.~\cite{gup-hug-pre-wis} in three euclidean dimensions in
section-II. In the secton-III  , we explicitly obtained the vertex
operator $ V\sim  e^{-i\,Q\,\phi}$ for the correlation function of
scalar field operators, $\phi \,$ (dual to electromagnetic field
strength). In the semiclassical
approximation , it is shown that at low energies these correlations
are reproduced in a flat background with the insertion of these vertex
operators $V \,$. One can interpretate that the insertion of vertex
operator in asymptotic flat space is equivalent to the effect of wormhole
background on the propagation of fields far away from the throat.
\par Next, in  section-IV we have considered the propagation of matter
fields in the wormholes geometry. We find that  vertex functions are
appropriately modified. The modification accounts for
introduction of field operators located at the throat of the
wormhole. These vertex functions,
when inserted in the low energy effective action, will reproduce
correctly the low energy Green functions for the Lorentz invariant
composite field operators made of scalar and fermi fields.
\par So far it has not been known to us how the above programme could
be extended for massive  ($m\ll a_o^{-1}$), or massless but not
conformaly coupled fields.

\acknowledgements
 I am deeply indebted to Prof. J. Maharana
for numerous invaluable discussions and for the suggessions while going through
the
reading of this manuscript.


\begin{references}
\bibitem[*]{mail} e-mail: hsingh@iopb.ernet.in

\bibitem{haw1} S. W. Hawking, Phys. Rev. D37(1988)904.\\
 See also I. Klebanov, L. Susskind and T. Banks, Nucl Phys B317(1989)665.

\bibitem{col2} S. Coleman, Nucl Phys B307(1988)867.\\
 S. B. Giddings and A. Strominger, Nucl Phys B307(1988)854.

\bibitem{col1} S. Coleman, Nucl Phys B310(1988)643.
\bibitem{pre} J. Preskill, Nucl Phys B323(1989)141.\\
 B. Grinstein and M. Wise, Phys Lett B212(1988)407.\\
  J. Preskill, S. Trivedi and Mark B. Wise, Phys Lett B223(1989)26.\\
  S. B. Giddings and A. Strominger, Phys. Lett. B230 (1989) 46.\\
 S. Coleman and K. Lee, Nucl. Phys. B329(1990) 387.\\
 C. P. Burgess and A. Kshirsagar, Nucl. Phys. B324 (1989) 157.\\
 S. Carlip and S. P. de Alwis, Nucl. Phys. B337 (1990) 681.\\
 W. Fischler and L. Susskind, Phys. Lett. B217 (1989) 48.\\
 W. Fischler, I. Klebanov, J. Polchinski and L. Susskind, Nucl. Phys.
B327 (1989) 157.

\bibitem{gup-hug-pre-wis} A. K. Gupta, J. Huge, J. Preskill and M.
                       Wise, Nucl Phys B333(1990)195.\\
                     A. Hosoya and W. Ogura, Phys Lett B225(1989)117.

\bibitem{kha-mah} S. P. Khastgir and J. Maharana, Phys. Lett. B301 (1993)
191.\\
				  S. P. Khastgir and J. Maharana, Nucl.	Phys. B406
(1993) 145.
 \bibitem{haw3} S. W. Hawking, Nucl Phys B335(1990)155.

\bibitem{gri-mah} B. Grinstein and J. Maharana, Nucl phys B333(1990)160.\\
                  B. Grinstein, J. Maharana and D. Sudarsky, Nucl Phys
B345(1990)231.

\bibitem{gid-str1} S. B. Giddings and A. Strominger, Nucl Phys B306(1988)890.

\bibitem{mye} R. C. Myers, Phys Rev D38(1988)1327.

\bibitem{gri} B. Grinstein, Nucl Phys B321(1989)439.

\bibitem{haw4} see S. W. Hawking in 'General Relativity', An Einstein
Centenary Survey, Cambridge Univ. Press (1979), eds. S. W. Hawking and W.
Israel.

\end{references}
\end{document}